\documentclass[aps,prev, superscriptaddress, preprintnumbers,floatset,nofootinbib,twocolumn]{revtex4-1}
\pdfoutput=1
\usepackage[english]{babel}
\usepackage{bm}
\usepackage{times}
\usepackage{ulem}
\usepackage{hyperref}
\hypersetup{
    colorlinks=true,
    linkcolor=blue,
    filecolor=magenta,      
    urlcolor=blue,
    citecolor=blue,
    pdftitle={Sharelatex Example},
    pdfpagemode=FullScreen
    }
\usepackage{listings}
\usepackage{slashed}
\usepackage{color}
\usepackage{appendix}
\usepackage{mathtools}
\usepackage{epsfig}
\usepackage{dcolumn}
\usepackage{textcomp}
\usepackage{appendix} 
\usepackage{multirow}
\usepackage{graphicx}
\usepackage{soul}
\usepackage{subfigure}
\usepackage[compat=1.1.0]{tikz-feynman} 
\newcommand{\fig}[1]{Fig.~\ref{#1}}

\begin{document} 

\title{The Hubble Tension: Relativistic Dark Matter Production from Long-lived Particles}

\author{Alvaro S. de Jesus}
\email{alvarosndj@gmail.com}
\affiliation{Departamento de F\'isica Matem\'atica, Instituto de F\'isica, Universidade de S\~{a}o Paulo, 05315-970, S\~ao Paulo, Brasil}

\author{Matheus M. A. Paixão}
\email[Corresponding author: ]{matheus.mapaixao@gmail.com}
\affiliation{Departamento de F\'isica, Universidade Federal do Rio Grande do Norte, 59078-970, Natal, RN, Brasil}
\affiliation{International Institute of Physics, Universidade Federal do Rio Grande do Norte,  59078-970, Natal, RN, Brasil}

\author{D\^eivid R. da Silva}
\email{deivid.rodrigo.ds@gmail.com}
\affiliation{Centro Brasileiro de Pesquisas Físicas, 22290-180, Rio de Janeiro, RJ, Brasil}

\author{Farinaldo S. Queiroz}
\email{farinaldo.queiroz@ufrn.br}
\affiliation{Departamento de F\'isica, Universidade Federal do Rio Grande do Norte, 59078-970, Natal, RN, Brasil}
\affiliation{International Institute of Physics, Universidade Federal do Rio Grande do Norte,  59078-970, Natal, RN, Brasil}
\affiliation{Millennium Institute for Subatomic Physics at High-Energy Frontier (SAPHIR), Fernandez Concha 700, Santiago, Chile}

\author{Nelson Pinto-Neto}
\email{nelsonpn@cbpf.br}
\affiliation{Centro Brasileiro de Pesquisas Físicas, 22290-180, Rio de Janeiro, RJ, Brasil}

\begin{abstract}

The tension between direct measurements of the Hubble constant and those stemming from Cosmic Microwave Background probes has triggered a multitude of studies. The connection between cosmology and particle physics has shown to be a valuable approach to addressing the Hubble tension. In particular, increasing the number of relativistic degrees of freedom in the early universe helps alleviate the problem. In this work, we write down effective field theory describing relativistic dark matter production in association with neutrinos leading to a larger $H_0$. We derive limits on the effective energy scale that governs this relativistic production of dark matter as a function of the dark matter mass for fermion, vector, and scalar dark matter fields. In particular, scalar dark matter particles are more effective in increasing the effective number of relativistic species. Also, if they have weak scale masses, then the relativistic production of dark matter should be governed by Planck scale effective operators in order to alleviate the Hubble tension.
\noindent

\end{abstract}


\maketitle
\flushbottom

\section{\label{In} Introduction}

The precise determination of the Hubble parameter, $H_0$, which quantifies the current expansion rate of the universe, has led to a notable discrepancy between values derived from late-time measurements and those obtained from early-universe probes—a conflict now widely referred to as the Hubble tension \cite{DiValentino2021,ref:Bernal2016,ref:Verde2019}. 

Late-time measurements, including those based on observations of Cepheid variable stars, Type Ia supernovae, and gravitational lensing, consistently yield a value around $73\,{\rm km s^{-1} Mpc^{-1}}$. However, the exact figure varies depending on the specific methods, datasets, and techniques employed in these observations.

For instance, the Supernovae and $H_0$ for the Equation of State (SH0ES) collaboration employs Type Ia supernovae as standard candles. By utilizing the local distance ladder method, which calibrates distances to nearby galaxies using Cepheid stars, the SH0ES team has determined an expansion rate of $73.2\pm1.3 \,{\rm km s^{-1} Mpc^{-1}}$ \cite{Riess:2020fzl}. Similarly, the $H_0$ Lenses in COSMOGRAIL's Wellspring (H0LiCOW) collaboration used gravitational lensing time delays, arriving at a value of $73.3\pm1.7 \,{\rm km s^{-1} Mpc^{-1}}$ \cite{Wong:2019kwg}. Additionally, recent observations of "standard sirens" by the LIGO/Virgo collaboration have provided an independent measurement of the Hubble constant using gravitational wave events \cite{LIGOScientific:2017adf,GW170817}. A combined analysis of signals from events GW190814, GW170814, and GW170817 yielded a Hubble parameter of $72.0^{+12}_{-8.2}\,{\rm km s^{-1} Mpc^{-1}}$ \cite{DES:2020nay}.

On the other hand, the values of $H_0$ obtained from early-time measurements based on the CMB power spectrum yield $H_0$ equal to $67.27 \pm 0.6 \,{\rm km s^{-1} Mpc^{-1}}$ for the $\Lambda$CDM model \cite{ref:Planck2018}, being statistically incompatible with the local measurements aforementioned, with some differences exceeding the $5\sigma$ threshold \cite{DiValentino2021, Riess:2020fzl}.

In order to understand such tension, we can explore two possible approaches. If we assume that the early-time measurements are indeed correct, we might attribute the higher values obtained by the local observations to a combination of systematic errors from different origins. In this case, unknown uncertainties related to Cepheids calibration, Type Ia supernovae modeling, and lens geometry could impact the accuracy of the results \cite{Freedman:2021ahq}. The perspective is that future observations may shed some light on this question by providing different manners to obtain $H_0$ and allowing comparisons with the current results \cite{DiValentino2021,Freedman:2021ahq}. Concerning gravitational-wave events, it is expected that as more standard sirens are observed, the associated error will be reduced, providing an independent and more competitive measurement of the Hubble parameter that can be used to validate or challenge the existing estimates \cite{Valentino2018,Feeney2019}. 

Additionally, measurements involving the Tip of Red Giants Branch (TRGB) provide an intermediate value for the Hubble constant, which is closer to the value coming from the CMB power spectrum analyses. The TRGB calibration applied to the Type Ia supernovae sample from the Carnegie Supernova Project (CSP) yields an $H_0$ of $69.8\pm0.6 \hspace{1mm}(\mathrm{sta})\pm1.6 \hspace{1mm} (\mathrm{sys}) \,{\rm km s^{-1} Mpc^{-1}}$ being consistent with SH0ES within $2\sigma$ \cite{Freedman:2021ahq}.

More recently, the James Webb Space Telescope (JWST) has provided high-resolution near-infrared data, addressing several systematic uncertainties associated with extragalactic measurements \cite{Freedman2024}. In relation to Cepheid variables, JWST is expected to mitigate the effects of crowding, refine constraints related to metallicity, and improve corrections for dust extinction. Moreover, JWST offers three independent methods to measure the distance to each galaxy: using Cepheid variables, Tip of the Red Giant Branch (TRGB) stars, and J-Region Asymptotic Giant Branch (JAGB) stars \cite{Freedman2023}. Recent observations that combine these methods have yielded a Hubble constant value of $69.96 \pm 1.05 , (\mathrm{stat}) \pm 1.12 , (\mathrm{sys}) \,{\rm km s^{-1} Mpc^{-1}}$ \cite{Freedman2024}, significantly reducing the Hubble tension.

Another possibility is to consider the late-time measurements as correct, in favor of a higher value of the Hubble parameter. In this case, it may imply that the $\Lambda$CDM model might be incomplete or require modifications to account for new physics that could yield a higher value of $H_0$ from the Planck data \cite{Riess2019,Riess2022}. Possible extensions to the model include an early dark energy stage \cite{Poulin2019}, non-standard Dark Matter \cite{Tulin2018}, self-interacting dark radiation \cite{Bagherian:2024obh} and relativistic production of dark matter can alleviate the $H_0$ tension \cite{ref:Hooper2011,ref:Kelso2013, Deivid:2022, Deivid:2023,daCosta2024}. In the latter, the increase in $H_0$ occurs via at the price of a larger $N_{eff}$.

Indeed, when considering a $\Lambda$CDM + $N_{eff}$ model, where $N_{eff}$ is allowed to vary, larger values of $H_0$ are favored (see Fig. \ref{fig:Neff_H0}). Specifically, using Baryon Acoustic Oscillations measurements derived from galaxy redshift surveys, Lyman-$\alpha$ data, and Big Bang Nucleosynthesis data on primordial abundances of helium and deuterium, it has been shown that increasing $N_{eff}$ by $0.1 < \Delta N_{eff} < 0.5$ can result in $H_0 > 67$ km/s/Mpc, thus alleviating the Hubble tension \cite{Schoneberg:2019wmt}.

 \begin{figure}[htb!]
    \centering
    \subfigure[]{
    \includegraphics[width=\columnwidth]{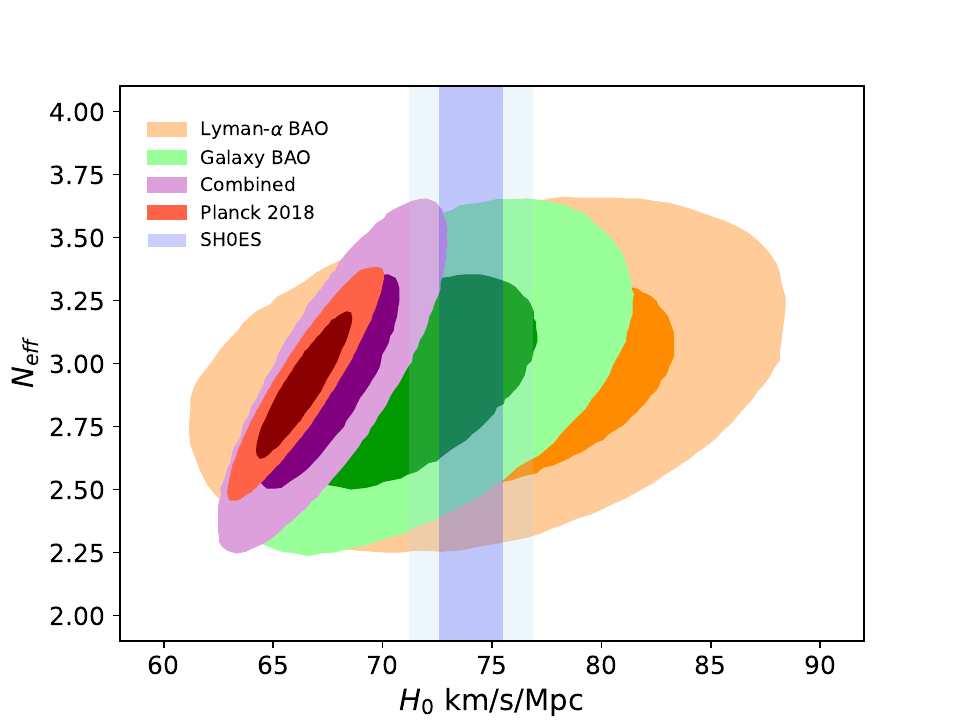}
    \label{fig:Neff_H0}}
    \caption{Relation between $N_{eff}$ and $H_0$ based on different datasets. The orange region reflects the BAO estimated from Lyman-$\alpha$, the green region the BAO inferred from galaxy surveys, the purple one is combined data. The red region is obtained using Planck CMB data only. The blue central area designates the observations from SH0ES. Figured adapted from \cite{Schoneberg:2019wmt}.}
    \label{fig:Neff_H0}
\end{figure}

Given the connection between $N_{eff}$ and $H_0$ illustrated in Fig. \ref{fig:Neff_H0}, we explore a scenario involving a long-lived heavy particle, $\chi'$, which thermally decoupled from the primordial plasma during the early stages of the universe and subsequently decayed into a dark matter particle $\chi$ and a massless neutrino $\nu$, via the process $\chi' \rightarrow \chi + \nu$. Assuming $m_{\chi'} \gg m_{\chi}$, the resulting dark matter particle $\chi$ would be relativistic at the time of production. The presence of this hot dark matter contributes additional radiation during the radiation-dominated era, effectively behaving like relativistic neutrinos and thus increasing the effective number of relativistic species, $N_{eff}$, and subsequently raising the value of $H_0$ \cite{mukhanov2005,dodelson2020,kolb2018early,ref:Anchordoqui2021, SunnyVagnozzi2020,Abdalla:2022yfr}. We model this relativistic production of dark matter using effective operators for scalar, fermion, and vector dark matter particles and delineate the parameter space in which this mechanism can achieve $H_0 > 67\,{\rm km s^{-1} Mpc^{-1}}$, thereby alleviating the Hubble tension. 

This article is structured as follows: In Section \ref{sec:sectionII} we review how this production mechanism can lead to a larger number of relativistic degrees of freedom. In Section \ref{sec: Effective models}, we write down effective operators and derive the connection with $N_{eff}$. In Section \ref{conclusions1} we draw our conclusions.

\section{Relativistic Production of Dark Matter and $N_{eff}$}\label{sec:sectionII}

\subsection{Defining $N_{eff}$}
In the $\Lambda$CDM model, the Hubble parameter can be expressed at early times as, 

\begin{equation}
    H^2=\frac{8\pi G} {3}\left(\rho_{\mathrm{mat}}+\rho_{\mathrm{rad}}\right).\label{FE}
\end{equation}

 We can decompose $\rho_{\mathrm{rad}}=\rho_{\gamma}+\rho_{1\nu}N_{\nu}$, with $\rho_{\gamma}$ being the energy density of the photons, and $\rho_{1\nu}$ accounting for the energy density of a single neutrino species in the Standard Model, where $N_{\nu}$ essentially represents the number of neutrino species.

Our hypothesis is that, before the matter-radiation equality, a long-lived particle $\chi'$ decays into two other particles: a dark matter particle $\chi$ and a massless neutrino via $\chi'\rightarrow\chi+\nu$. If the $\chi'$ particle is much heavier than $\chi$, $m_{\chi'}\gg m_{\chi}$, the resulting dark matter particle will be relativistic at the moment of its production. Therefore, it does not contribute to $\rho_{\mathrm{mat}}$, but actually to $\rho_{\mathrm{rad}}$, acting as an extra source of radiation. As a consequence, 

\begin{equation}
    \rho_{\mathrm{rad}}=\rho_{\gamma}+\rho_{1\nu}N_{\nu}+\rho_{\mathrm{extra}},
\end{equation}
from which we can see that this dark radiation will mimic the effect of an extra neutrino. For this purpose, we just need to express the extra energy as, 
\begin{equation}
    \rho_{\mathrm{extra}}=\rho_{1\nu}\Delta N_{eff},
\end{equation}
with $\Delta N_{eff}$ being this extra number of relativistic species. Defining $N_{eff}=N_{\nu}+\Delta N_{eff}$ we obtain,
\begin{equation}
    \rho_{\mathrm{rad}}=\rho_{\gamma}+\rho_{1\nu}N_{eff},
\end{equation}
with $N_{eff}$ being larger than 3. This is precisely the additional parameter that is usually incorporated into CMB analyses and that raises the value of $H_0$ \cite{DiValentino2021}.

As we wish to reconcile early and late time determinations of $H_0$ through $N_{eff}$ we compute $\Delta N_{eff}$ at the matter-radiation equality. That said, we define $\Delta N_{eff}$ as,
\begin{equation}
    \Delta N_{eff}=\left.\frac{\rho_{\mathrm{extra}}}{\rho_{1\nu}}\right|_{t=t_{\mathrm{eq}}},\label{DeltaN}
\end{equation}
where $t_{\mathrm{eq}}$ designs the time of matter-radiation equality, what happens approximately 50.000 years after the Big Bang \cite{dodelson2020}. 

Interpreting $N_{eff}$ as an additional neutrino family conflicts with cosmological data  \cite{DiValentino2021} and accelerator measurements \cite{ParticleDataGroup:2002ivw}. In our approach the contribution to $N_{eff}$ does not come from a new neutrino species, but rather from relativistic production of dark matter particles which eventually become non-relativistic. This brief relativistic behavior mimics an extra neutrino species at the matter-radiation equality. 

From the BBN, $\Delta N_{eff}<0.5$ at 95\% of confidence level \cite{Knapen2017,Cyburt2016,Nollett2015}. Assuming the $\Lambda$CDM model, the CMB bound is even more restrictive, pointing to $\Delta N_{eff}<0.3$ from Planck TT, TE, EE + lowE + lensing + BAO dataset \cite{ref:Planck2018}. A similar limit was obtained taking into account the additional effects of curvature, dark energy, massive neutrinos, and weak scale-dependence, considering the results provided by Planck with respect to the temperature and polarization likelihood \cite{ref:Planck2018,Planck2018I,Planck2018V,Valentino2020,Valentino2022,Yang2023}. On top of that, the South Pole Telescope polarization measurements \cite{Dutcher2021} together with the 9 years results from WMAP observations \cite{Hinshaw2013} provided a superior bound on $\Delta N_{eff}$ approaching the value indicated by BBN \cite{Valentino2022}. 

In what follows, we will not derive the connection between the relativistic production of dark matter with $N_{eff}$.

\subsection{Relativistic Production}

Considering that the long-lived particle decays at rest, the momenta of the particles participating in the decay $\chi'\rightarrow\chi+\nu$ are,
\begin{align}
    &p_{\chi'} = (m_{\chi^\prime}, \bm{0}),\nonumber\\
    &p_{\chi} = (E_{\chi}(\bm{p}), \bm{p}),\\
    &p_{\nu} = (|\bm{p}|, -\bm{p})\nonumber,
\end{align}
where $E_{\chi}$ and $\bm{p}$ are respectively the energy and the tri-momentum associated with the dark matter particle $\chi$, being related with the masses $m_{\chi}$ and $m_{\chi'}$ via,

\begin{align}
    E_{\chi}(\tau) &= m_{\chi} \left( \frac{m_{\chi^\prime}}{2m_{\chi}} + \frac{m_{\chi}}{2m_{\chi^\prime}} \right), \label{eq:energy_in_tau}\\
    \left|\bm{p}_{\chi}(\tau) \right| &= \left|\bm{p}\right| = \frac{1}{2} m_{\chi^\prime} \left[ 1 - \left(\frac{m_{\chi}}{m_{\chi^\prime}} \right)^2 \right]. \label{eq:momentum_chi}
\end{align}
Here, we suppose that all the decays occur at $t=\tau$, with $\tau$ being the lifetime of $\chi'$. Please note that the energy equation can be written as $E_{\chi} (\tau)=m_{\chi}\gamma_{\chi} (\tau)$ with,
\begin{equation}
    \gamma_{\chi}(\tau)=\frac{m_{\chi^\prime}}{2m_{\chi}} + \frac{m_{\chi}}{2m_{\chi^\prime}}
\end{equation}
being the Lorentz factor. Taking into account the expansion history we can express the energy of the dark matter particle as,
\begin{equation}
    E_{\chi}(t) = m_{\chi}\left[ 1 + \left( \frac{a(\tau)}{a(t)} \right)^2 \left( \gamma^2_{\chi}(\tau) - 1\right)\right]^{1/2}.
\end{equation}
As the universe was radiation-dominated at the time of decay, the ratio between $a(\tau)$ and $a(t)$ in the last equation is just $\sqrt{\tau/t}$, which implies that,
\begin{equation}
    \gamma_{\chi}(t) = \left\{1+\frac{ (m^2_{\chi'} - m^2_{\chi})^2 }{4m^2_{\chi}m^2_{\chi'}} \left( \frac{\tau}{t} \right) \right\}^{1/2}.\label{BoostFactor}
\end{equation}

Eq.\eqref{BoostFactor} determined the boost factor experienced by the dark matter species produced from the decay of the long-lived particle $\chi^\prime$. As we are assuming $m_{\chi^\prime} \gg m_{\chi}$, $\gamma_\chi > 1$. It is well-known that the bulk of dark matter should not be relativistic at the matter-radiation equality. Therefore, we assume that only a fraction, $f$, of the overall dark matter density is produced via this mechanism. In other words, our mechanism relies on the existence of hot and cold dark matter components. The latter being produced in some way, and former via our mechanism.  

\subsection{$N_{eff}$ from long-lived decays}

In order to separate the cold and hot  dark matter components we can write the energy equation of all the dark matter as,
\begin{equation}
    E_{\chi}=m_{\chi}N_{\mathrm{CDM}}+(\gamma_{\chi}-1)m_{\chi}N_{\mathrm{HDM}},
\end{equation}
where $N_{\mathrm{CDM}}$ and $N_{\mathrm{HDM}}$ are the numbers of the non-relativistic and relativistic dark matter particles, respectively, with $N_{\mathrm{HDM}}\ll N_{\mathrm{CDM}}$ so that we do not violate any bounds imposed by Big Bang Nucleosynthesis or structure formation. As a consequence, the ratio between the energy densities of these two dark matter components is 
\begin{equation}
    \frac{\rho_{\mathrm{HDM}}}{\rho_{\mathrm{CDM}}} = \left( \gamma_{\chi} -1 \right)f,
\end{equation}
where the factor $f$ is the ratio between the number densities of the hot and cold dark matter constituents.

Considering that the extra source of radiation outlined in Eq. \eqref{DeltaN} is completely due to the decay, then
\begin{equation}
    \Delta N_{eff}=\left.\frac{\rho_{\mathrm{CDM}}}{\rho_{1\nu}}\right|_{t=t_{\mathrm{eq}}}(\gamma_{\chi}(t_{\mathrm{eq}})-1)f.
\end{equation}
Since at matter-radiation equality $\rho_{\mathrm{CDM}}=\rho_{\mathrm{crit}}\Omega_{\mathrm{CDM}}a_{\mathrm{eq}}^{-3}$ and $\rho_{1\nu}=\rho_{\mathrm{crit}}\Omega_{\nu}a_{\mathrm{eq}}^{-4}/3$, where $\rho_{\mathrm{crit}}$ is the critical density of the universe, and $\Omega_{\mathrm{CDM}}$ and $\Omega_{\nu}$ are the density parameters obtained by Planck for (cold) dark matter and neutrinos, respectively, it follows that the energy density associated with one neutrino species is 16\% of the (cold) dark matter density \cite{ref:Planck2018}. Thus,
\begin{equation}
    \left.\dfrac{\rho_{1\nu}}{\rho_{\mathrm{CDM}}}\right|_{t=t_{\mathrm{eq}}}=0.16.
\end{equation}

In the limit of $m_{\chi'}\gg m_{\chi}$, which is necessary for $\chi$ to be relativistic, we can express $\Delta N_{eff}$ in terms of the $\chi'$ lifetime and the mother-to-daughter mass ratio as,
\begin{equation}
        \Delta N_{eff} \approx 2.5 \times 10^{-3}\sqrt{\frac{\tau}{10^{6} \mathrm{s}}} \times f\frac{m_{\chi'}}{m_{\chi}}.
    \label{eq:deltaN}
\end{equation}

Using Eq.\eqref{eq:deltaN} we can relate particle physics quantities to the $N_{eff}$ and consequently to $H_0$. The key quantities are the lifetime of the long-lived particle and the $f \, m_{\chi'}/m_{\chi}$. Having described the mechanism in more detail, we address some cosmological bounds below.

\subsection{Cosmological bounds}

We will consider the constraints imposed by Structure Formation, Big Bang Nucleosynthesis (BBN), and Cosmic Microwave Background (CMB) observations. A more detailed discussion on the role of entropy injection is provided in Appendix \ref{secentropy}.

\paragraph{Structure Formation:} For the bottom-up structure formation process to occur, the majority of dark matter must be non-relativistic at the time of matter-radiation equality. Structure formation models constrain the relativistic component of dark matter to be at most 6\% of the total dark matter density when structures begin to form  \cite{ref:Kelso2013, ref:Alcaniz2021,Allahverdi2014,Zhao}. Therefore, we conservatively assume that only 1\% of the total dark matter budget is produced relativistically through this mechanism, i.e., $f=0.01$. We remain agnostic about the primary dark matter production mechanism responsible for the non-relativistic component \cite{Arcadi:2024ukq}.

\paragraph{Big Bang Nucleosynthesis:}
The existence of long-lived particles with lifetime $\tau> 1$~s may induce electromagnetic injection into the plasma and alter the abundance of light elements \cite{Alves:2023jlo}. Constraints from  Big Bang Nucleosynthesis limit us to the parameter space where $10^2$s $<\tau < 10^ 4$s and $10^3<m_{\chi'}/m_{\chi}<10^6$ \cite{Alves:2023jlo}. That said, we will restrict our analysis to this parameter space while producing a $\Delta N_{eff}>0.1$ and consequently a larger value of $H_0$. 
\paragraph{Cosmic Microwave Background:}

This relativistic production of dark matter species has been contemplated in the context of CMB. Using the CAMB code \cite{Lewis:1999bs}, a Monte Carlo simulation has yielded $H_0\approx 70\, {\rm km s^{-1}Mpc^{-1}}$ for $N_{eff}=3.3$ for both null and non-null curvatures \cite{Deivid:2023,daCosta2024} in agreement with the local measurements made by the James Webb Space Telescope \cite{Freedman2024}. Interestingly, the scenarios that reproduce  $H_0\approx 70\, {\rm km s^{-1}Mpc^{-1}}$ are consistent with the Big Bang Nuclesynthesis favored region. That said, we will target this parameter space in the next section, where we describe this relativistic production of dark matter using effective operators.

\section{Effective theory models}\label{sec: Effective models}

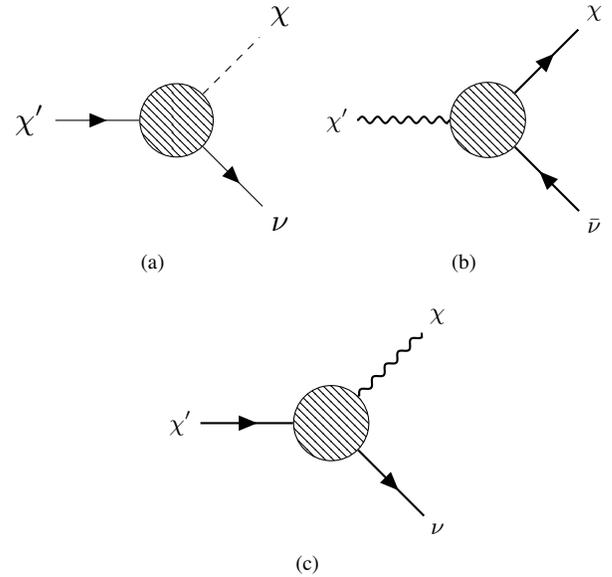
\begin{figure}[htb!]
    \centering
    \subfigure[]{
        \begin{tikzpicture}
            \begin{feynman}[scale = 1.3, transform shape]
                \vertex[blob] (a) {};
                \vertex [left = of a] (b) {$\chi^\prime$};
                \vertex [above right = of a] (c) {$\chi$};
                \vertex [below right = of a] (d) {$\nu$};
                \vertex [right = of a] (e);
                \diagram{
                    (b) --[fermion] (a) --[scalar] (c);
                    (a) --[fermion] (d);
                };
            \end{feynman}
        \end{tikzpicture}
        \label{fig:diagram4_FSN}
    }
    \subfigure[]{
        \begin{tikzpicture}
            \begin{feynman}[large]
                \vertex[blob] (a) {};
                \vertex [left = of a] (b) {$\chi^\prime$};
                \vertex [above right = of a] (c) {$\chi$};
                \vertex [below right = of a] (d) {$\bar{\nu}$};
                \vertex [right = of a] (e);
                \diagram{
                    (b) --[boson] (a) --[fermion] (c);
                    (d) --[fermion] (a);
                };
            \end{feynman}
        \end{tikzpicture}
        \label{fig:diagram5_VFN}
    }
    \subfigure[]{
        \begin{tikzpicture}
            \begin{feynman}[large]
                \vertex[blob] (a) {};
                \vertex [left = of a] (b) {$\chi^\prime$};
                \vertex [above right = of a] (c) {$\chi$};
                \vertex [below right = of a] (d) {$\nu$};
                \vertex [right = of a] (e);
                \diagram{
                    (b) --[fermion] (a) --[boson] (c);
                    (a) --[fermion] (d);
                };
            \end{feynman}
        \end{tikzpicture}
        \label{fig:diagram6_FVN}
    }
    \caption{Diagrammatic representation of a heavy particle $(\chi^\prime)$ that decay in hot dark matter $(\chi)$ and neutrino $(\nu)$. Three cases are considered: \textbf{(a)} $\chi^\prime$ is spin-1/2 and $\chi$ is a spin-0 particle; \textbf{(b)} $\chi^\prime$ has spin-1 and $\chi$ has spin-1/2 particles; \textbf{(c)} $\chi^\prime$ is spin-1/2 and $\chi$ is a spin-1 particle.}
    \label{fig:diagrams_neutrinos}
\end{figure}

Up until now, we have exploited the $N_{eff}-H_0$ relation in a more general setting. In this section, we will calculate the physical implications of the decay $\chi^\prime \rightarrow \chi + \nu$ for three distinct setups: (a) $\chi^\prime$ is a spin-1/2 particle and $\chi$ is a spin-0 particle; (b) $\chi^\prime$ has spin-1, and $\chi$ has spin-1/2; (c) $\chi^\prime$ is a spin-1/2 fermion, and $\chi$ is a spin-1 boson. In this way, we will cover fermion, scalar and vector dark matter production. All these cases are depicted in \fig{fig:diagrams_neutrinos}.

For each case, we write down n effective Lagrangian and then calculate the decay rate $\Gamma$. For a two-body decay process, the decay rate can be analytically calculated \cite{griffiths2020introduction} with,
\begin{equation}
    \Gamma(\chi^\prime \rightarrow \chi + \nu) = \frac{\left|\bm{p}_{\chi} (\tau) \right|}{8\pi m^2_{\chi^\prime}} \left| \mathcal{M} \right|^2,
    \label{eqGamma}
\end{equation}where $\mathcal{M}$ is the invariant amplitude of the simplified lagrangian.

We have already used the decay kinematics to determine the momentum of the $\chi$ particle at $t = \tau$, as seen in Eq. \ref{eq:momentum_chi}. Applying it to Eq.\eqref{eqGamma} we get, 
\begin{equation}
    \Gamma = \frac{1}{16\pi m_{\chi^\prime}}\left[ 1 - \left(\frac{m_{\chi}}{m_{\chi^\prime}} \right)^2 \right] \left| \mathcal{M} \right|^2.
    \label{eq:gamma_general}
\end{equation}

With this information, we are able to calculate the lifetime $\tau = 1/\Gamma$ for fermion, scalar, and vector dark matter production and feed that into the Eq.\eqref{eq:deltaN}. 

\subsection{Scalar Dark Matter}

Assuming $\chi^\prime$ is a fermion, $\chi$ a real scalar field, the  $\chi^\prime \rightarrow \chi + \nu$ decay is described by the effective lagrangian, 
\begin{equation}
    \begin{split}
            \mathcal{L}_{eff} & = \sum_{\mathrm{\nu = \nu_e, \nu_\mu, \nu_\tau}} \left[ \frac{1}{\Lambda_1}\bar{\psi}_{\nu} \frac{1}{2} \left(I + \gamma^5 \right) \gamma^\mu (\partial_\mu \psi_{\chi^\prime})\phi_{\chi} \right. \\ 
            & + \left. \frac{1}{\Lambda_2}\bar{\psi}_{\nu} \frac{1}{2} \left(I + \gamma^5 \right) \gamma^\mu \psi_{\chi^\prime}\partial_\mu\phi_{\chi} \right] + h.c. ,
    \end{split}
\end{equation}
where $\Lambda_1$ and $\Lambda_2$ are effective energy scales. The Feynman diagram for this process is shown in \ref{fig:diagram4_FSN}. Thus, the Feynman amplitude has three contributions, one for each neutrino, and its expression is,

\begin{equation}
    \sum_{\mathrm{\nu = \nu_e, \nu_\mu, \nu_\tau}} \left| \mathcal{M}\right|^2 = \frac{3m^4_{\chi^\prime}}{2} \left( \frac{1}{\Lambda_1} - \frac{1}{\Lambda_2} \right)^2 \left[ 1 - \left( \frac{m_{\chi}}{m_{\chi^\prime}} \right)^2 \right].\label{amplitude-case-A}
\end{equation}

Applying this result into Eq. (\ref{eq:gamma_general}) we find that,
\begin{eqnarray}
    \Gamma &=& \frac{3m^3_{\chi^\prime}}{32\pi}\left( \frac{1}{\Lambda_1} - \frac{1}{\Lambda_2} \right)^2 \left[ 1 - \left(\frac{m_{\chi}}{m_{\chi^\prime}} \right)^2 \right]^2, \nonumber\\
    \Rightarrow \Gamma &\approx& \frac{3m^3_{\chi^\prime}}{32\pi}\left( \frac{1}{\Lambda_1} - \frac{1}{\Lambda_2} \right)^2. \label{eq:gamma_FSN}
\end{eqnarray}

\begin{figure}[htb!]
    \centering
    \subfigure[]{
    \includegraphics[width=\columnwidth]{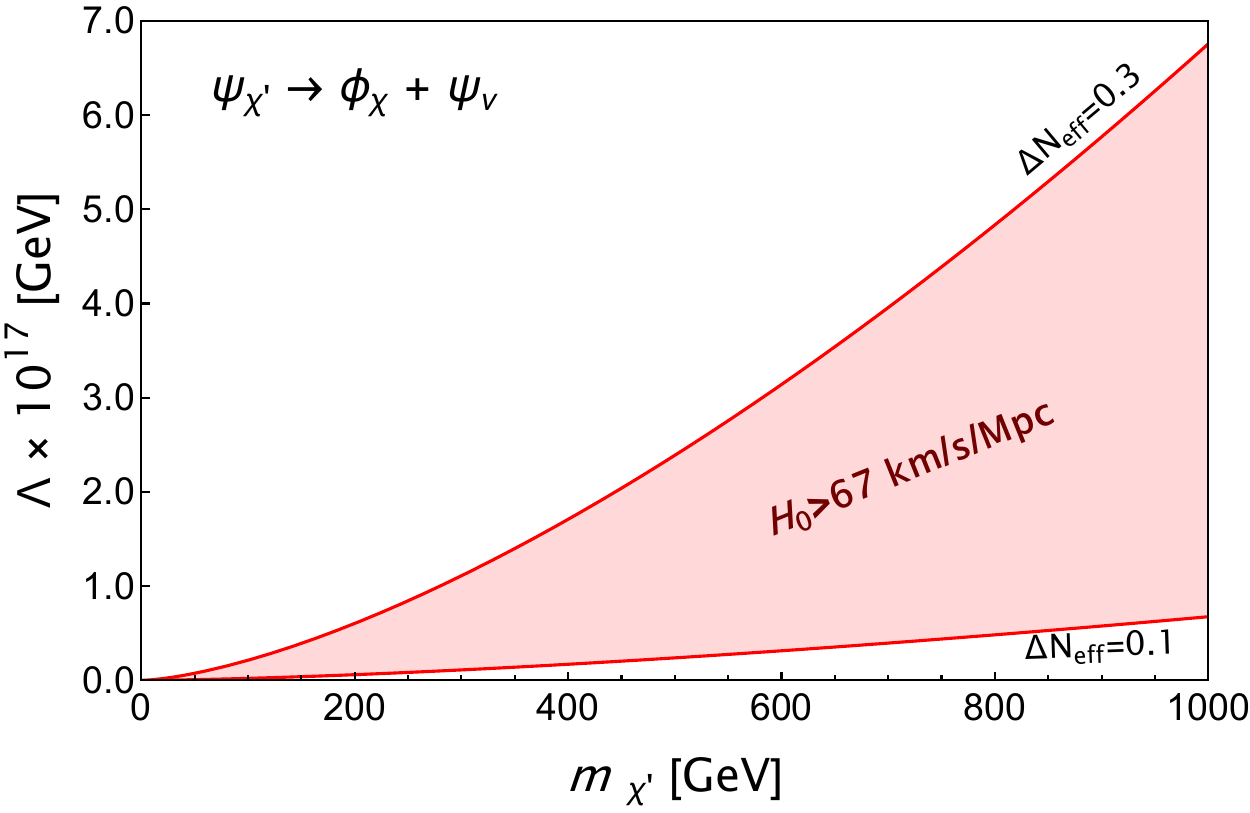}
    \label{fig:MxLambda_case4_FSN}}
    \subfigure[]{
    \includegraphics[width=\columnwidth]{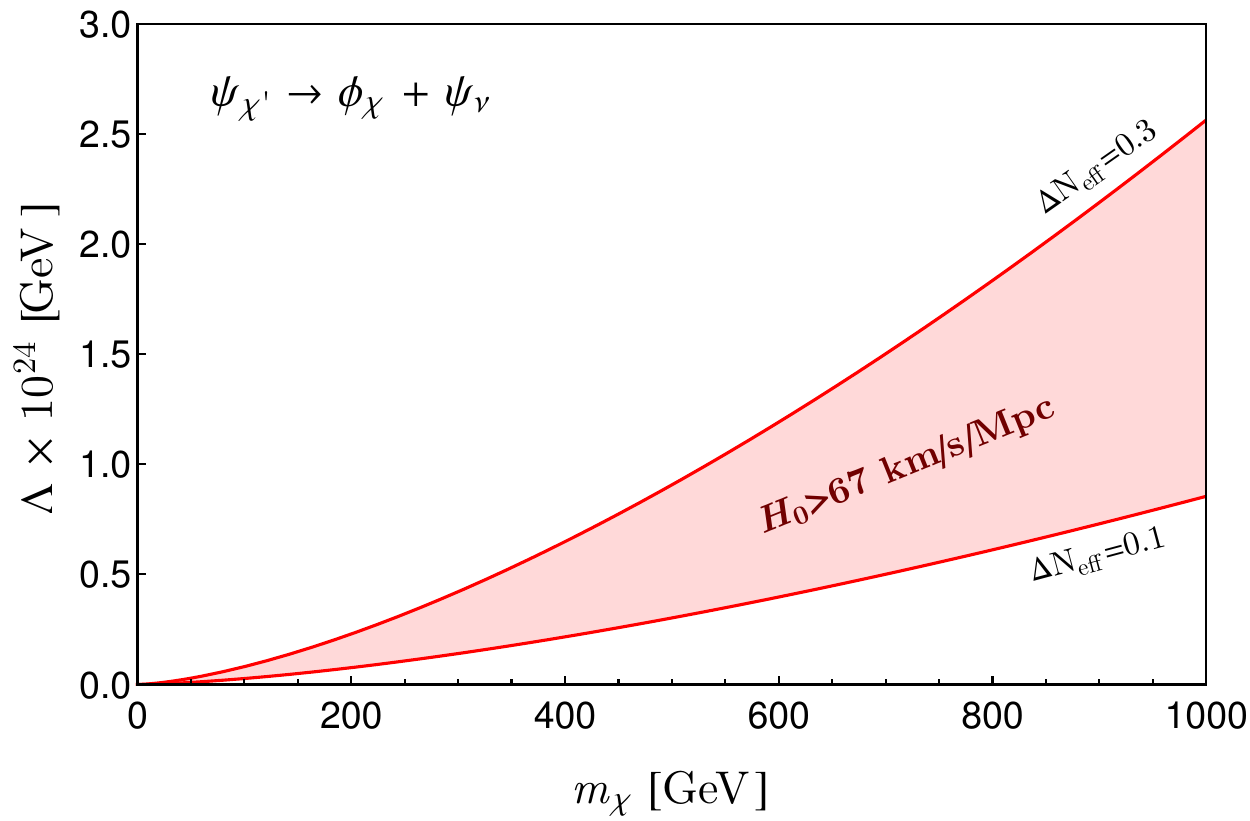}
    \label{fig:plot_Lambda_mchi_mod4}}
    \caption{We plot the energy scale $\Lambda$ as a function of $\chi'$ and $\chi$ masses for the case where $\chi^\prime$ is a fermion and $\chi$ a scalar. In the panel \textbf{(a)} we exhibit $\Lambda \times m_{\chi^\prime}$ delimited by $0.1 < \Delta N_{eff} < 0.3$. \textbf{(b)} we exhibit $\Lambda \times m_{\chi}$ instead. In both cases we adopted $f=0.01$ and $m_{\chi^\prime}/m_{\chi} = 10^3$.}
    \label{fig:plots_caseIV}
\end{figure}

From Eq.\eqref{eq:gamma_FSN} we conclude that for $\Lambda_1=\Lambda_2$ $\chi^\prime$ does not decay, and that the decay rate is governed by the smallest energy scale. Without loss of generality, taking $\Lambda_1=\Lambda \ll \Lambda_2$ we outline in \fig{fig:plots_caseIV} the region of parameter space that could alleviate the Hubble tension in the $\Lambda-m_{\chi^\prime}$ plane. Having in mind the aforementioned bounds, we fixed $f=0.01$ and $m_{\chi^\prime}/m_{\chi} = 10^3$, and allowed $10^2\text{s} < \tau < 10^4\text{s}$. The upper and lower curves that delimit the red region in Fig.\ref{fig:plots_caseIV} yield $\Delta N_{eff}=0.3$ and $\Delta N_{eff}=0.1$, respectively.

In Fig.\ref{fig:MxLambda_case4_FSN}, we conclude that a long-lived particle with a lifetime around $10^2-10^4$~s with masses at the weak scale can reduce the $H_0$ tension for an effective energy scale around $10^{17}$~GeV. We remind the reader that we impose $m_{\chi^\prime} \gg m_\chi$ throughout. Therefore, the dark matter masses should be below $1$~GeV in this setup. If we ramp up the dark matter mass to be at the weak scale instead, the effective energy scale that drives the $\chi^\prime$ decay goes to the Planck scale as can be seen in \ref{fig:plot_Lambda_mchi_mod4}. 

The relationship between Eq. \eqref{eq:gamma_FSN} and $\Delta N_{eff}$ (Eq.\eqref{eq:deltaN}) is straightforward:
\begin{equation}
    \Delta N_{eff} = 1.17 \times 10^{-17}\frac{\Lambda}{m_{\chi}}\sqrt{\frac{\mathrm{GeV}}{m_{\chi^\prime}}}f.
    \label{eq:deltaNeff_caseA}
\end{equation}

With this result, it is clear that $\Delta N_{eff}$ depends only on the particle physics properties. Consequently, this also holds for the increase in $H_0$.

\subsection{Fermion Dark Matter}

Considering $\chi^\prime$ to be vector boson and $\chi$ a fermion the simplified lagrangian that accounts for the  $\chi^\prime \rightarrow \chi + \bar{\nu}$ decay reads, 
\begin{equation}
    \mathcal{L}_{eff} = \sum_{\nu = \nu_e, \nu_\mu, \nu_\tau} \frac{1}{\Lambda}\bar{\psi}_\chi \sigma^{\mu \nu}\chi^\prime_{\mu \nu} \frac{1}{2} \left(I - \gamma^5 \right)\psi_{\nu} + h.c.,
\end{equation}
where $\chi^\prime_{\mu \nu} \equiv \partial_\mu \chi^\prime_{\nu} - \partial_\nu \chi^\prime_{\mu}$, and $\chi^\prime_\mu$ is the vector field that describe $\chi^\prime$. The Feynman amplitude for this model is
\begin{equation}
   \sum_{\nu = \nu_e, \nu_\mu, \nu_\tau} \left| \mathcal{M}\right|^2 = \frac{4 m_{\chi^\prime}^4}{\Lambda^2}\left[1+\left(\frac{m_\chi}{m_{\chi^\prime}}\right)^2 - 2\left(\frac{m_\chi}{m_{\chi^\prime}}\right)^4 \right].
   \label{amplitude-case-B}
\end{equation}

Plugging Eq.\eqref{amplitude-case-B} into 
Eq.\eqref{eq:gamma_general} we get,
\[\Gamma = \frac{m_{\chi^\prime}^3}{4\pi \Lambda^2}\left[ 1 - \left(\frac{m_{\chi}}{m_{\chi^\prime}} \right)^2 \right]\left[1+\left(\frac{m_\chi}{m_{\chi^\prime}}\right)^2 - 2\left(\frac{m_\chi}{m_{\chi^\prime}}\right)^4\right],\]which in the limit $m_{\chi^\prime} \gg m_{\chi}$ simplifies to,
\begin{equation}
    \Gamma \approx \frac{m_{\chi^\prime}^3}{4\pi \Lambda^2}\cdot \label{eq:gamma_VFN}
\end{equation}

This result is similar to the previous case up to some constant factors.  In a similar vein to the previous calculations we find, 

\begin{figure}[htb!]
    \centering
    \subfigure[]{
    \includegraphics[width=\columnwidth]{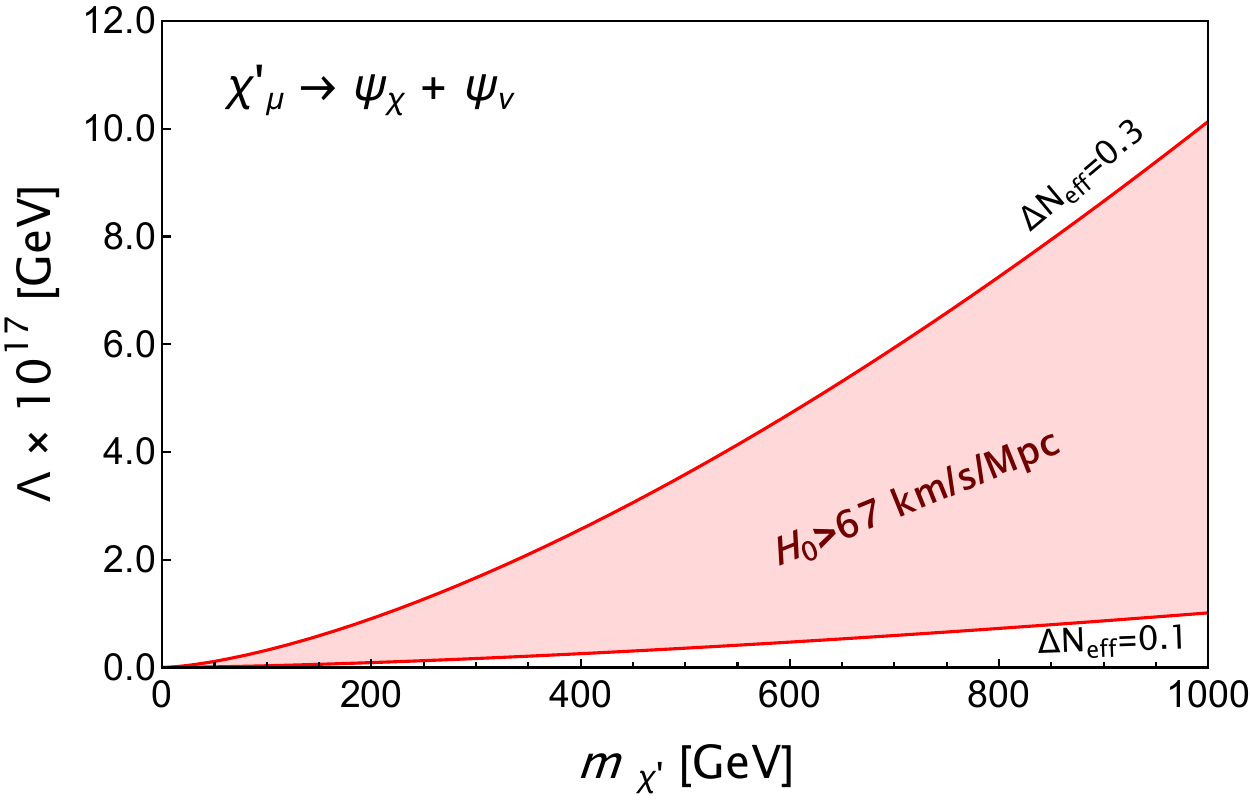}
    \label{fig:MxLambda_case5_VFN}}
    \subfigure[]{
    \includegraphics[width=\columnwidth]{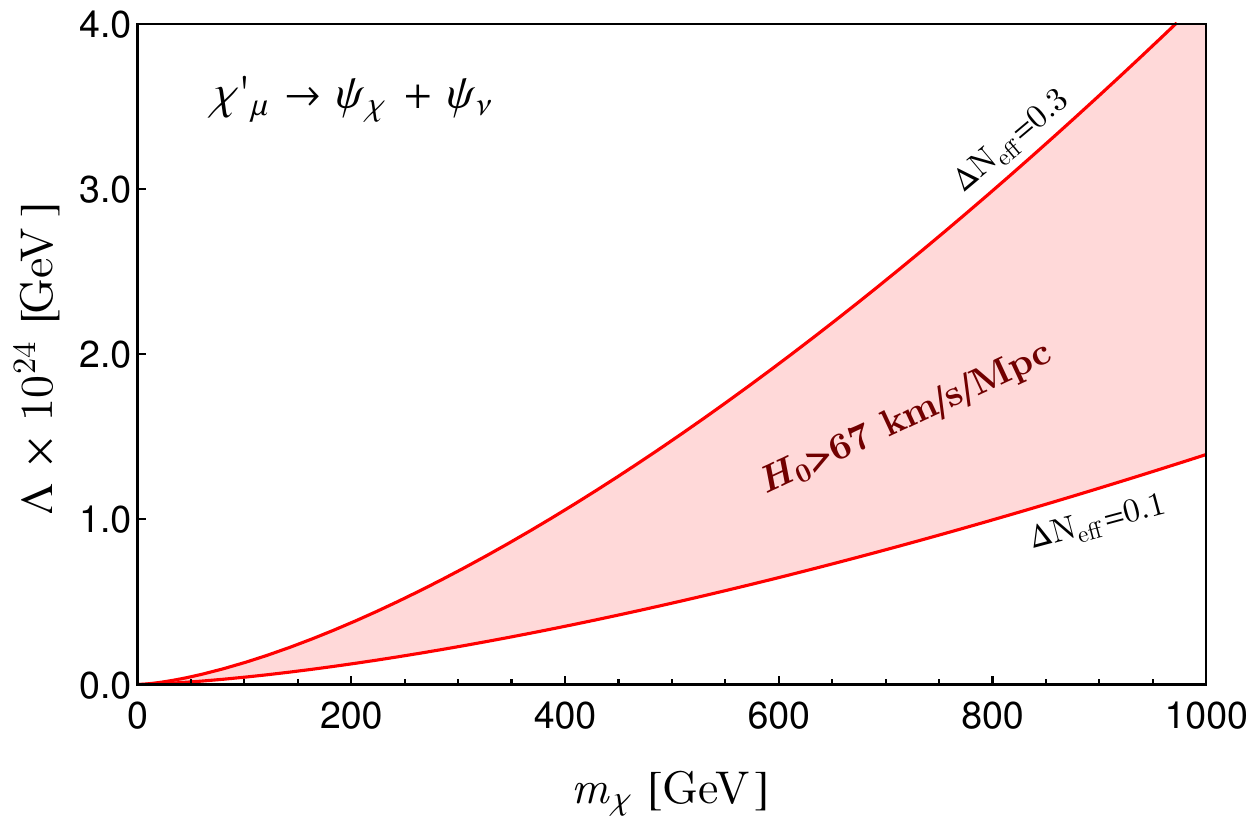}
    \label{fig:plot_Lambda_mchi_mod5}}
    \caption{Plot of $\Lambda$ versus $m_{\chi'}$ and $\Lambda$ versus $m_{\chi}$ for the case where $\chi^\prime$ has spin-1 and $\chi$ is a spin-1/2 dark matter particle. \textbf{(a)} $\Lambda \times m_{\chi^\prime}$ curves  built within the range $0.1 < \Delta N_{eff} < 0.3$. \textbf{(b)} Analogous plot, but considering $m_{\chi}$. The region of interest, marked by the hatched area, addresses the Hubble tension. We assume $f=0.01$ and $m_{\chi^\prime}/m_{\chi} = 10^3$ in both cases.}
    \label{fig:plots_caseV}
\end{figure}

\begin{equation}
    \Delta N_{eff} = 7.18\times10^{-18}\frac{\Lambda}{m_{\chi}}\sqrt{\frac{\mathrm{GeV}}{m_{\chi^\prime}}}f.
    \label{eq:deltaNeff_caseB}
\end{equation}

As the decay widths for scalar and fermion dark matter are equivalent up to some constant factors, the final plots of $\Lambda$ as a function of the particle masses are similar (See Fig.\ref{fig:MxLambda_case5_VFN} and Fig.\ref{fig:plot_Lambda_mchi_mod5}).

\subsection{Vector Dark Matter}

The last setup considered is vector dark matter. The $\chi^\prime$ field is a fermion while $\chi$ is a vector  particle that interact with neutrinos via the Lagrangian,
\begin{equation}
    \mathcal{L}_{eff} = \sum_{\nu = \nu_e, \nu_\mu, \nu_\tau} \frac{1}{\Lambda} \bar{\psi}_\nu \frac{1}{2} \left(I + \gamma^5 \right) \sigma^{\mu \nu}\psi_{\chi^\prime}\chi_{\mu \nu} + h.c.,
\end{equation}
where $\Lambda$ is a constant with dimension of mass, $\chi_{\mu \nu} = \partial_\mu \chi_\nu - \partial_\nu \chi_\mu$, and $\chi_\mu$ is the vector field that describe $\chi$. The Feynman amplitude for this model is,
\begin{equation}
    \sum_{\nu = \nu_e, \nu_\mu, \nu_\tau} \left| \mathcal{M}\right|^2 = \frac{6 m_{\chi^\prime}^4}{\Lambda^2}\left[2-\left(\frac{m_\chi}{m_{\chi^\prime}}\right)^2 - \left(\frac{m_\chi}{m_{\chi^\prime}}\right)^4\right] \label{amplitude-case-C}.
\end{equation}

Hence, the interaction rate is given by,

\[\Gamma = \frac{3 m_{\chi^\prime}^3}{8 \pi \Lambda^2}\left[ 1 - \left(\frac{m_{\chi}}{m_{\chi^\prime}} \right)^2 \right] \left[2-\left(\frac{m_\chi}{m_{\chi^\prime}}\right)^2 - \left(\frac{m_\chi}{m_{\chi^\prime}}\right)^4\right], \]which in the $m_{\chi^\prime} \gg m_{\chi}$ limit becomes,
\begin{equation}
    \Gamma \approx \frac{3 m_{\chi^\prime}^3}{4 \pi \Lambda^2} \cdot
    \label{eq:gamma_FVN}
\end{equation}

\begin{figure}[htb!]
    \centering
    \subfigure[]{
    \includegraphics[width=\columnwidth]{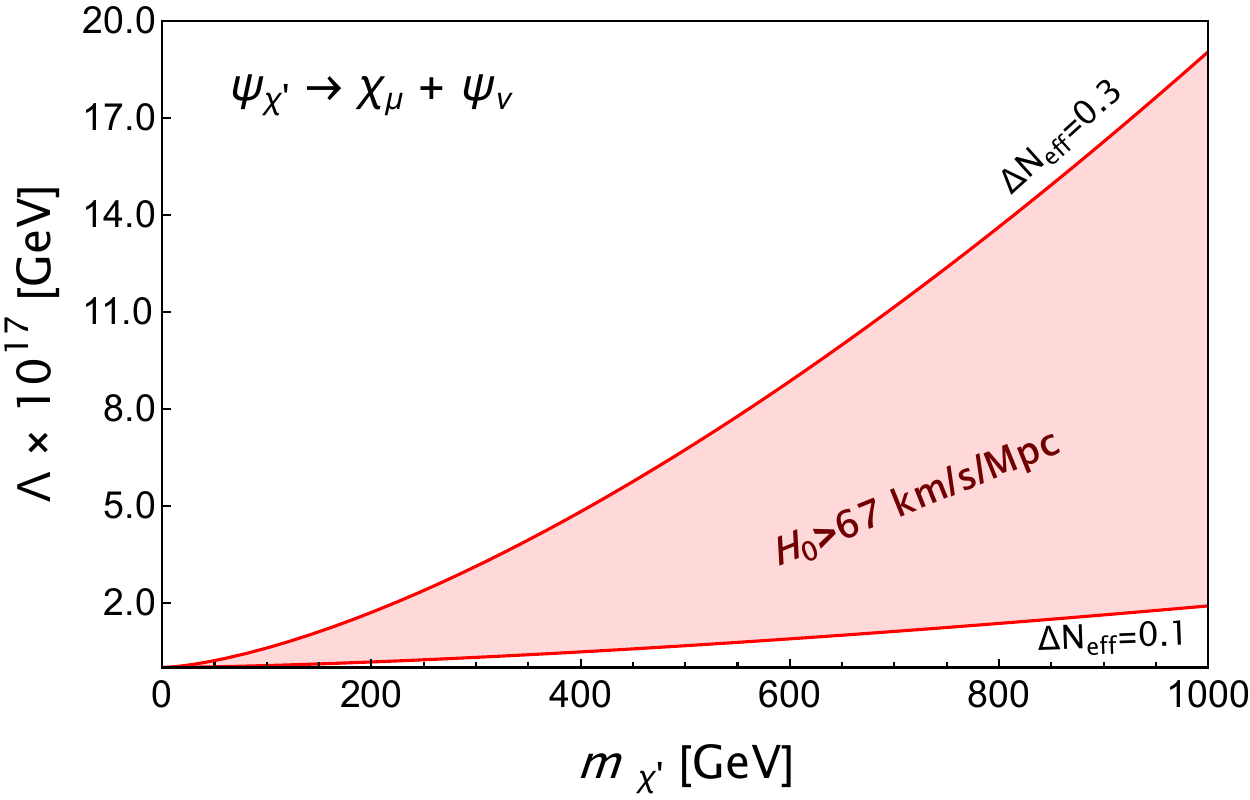}    \label{fig:MxLambda_case6_FVN}}
    \subfigure[]{
    \includegraphics[width=\columnwidth]{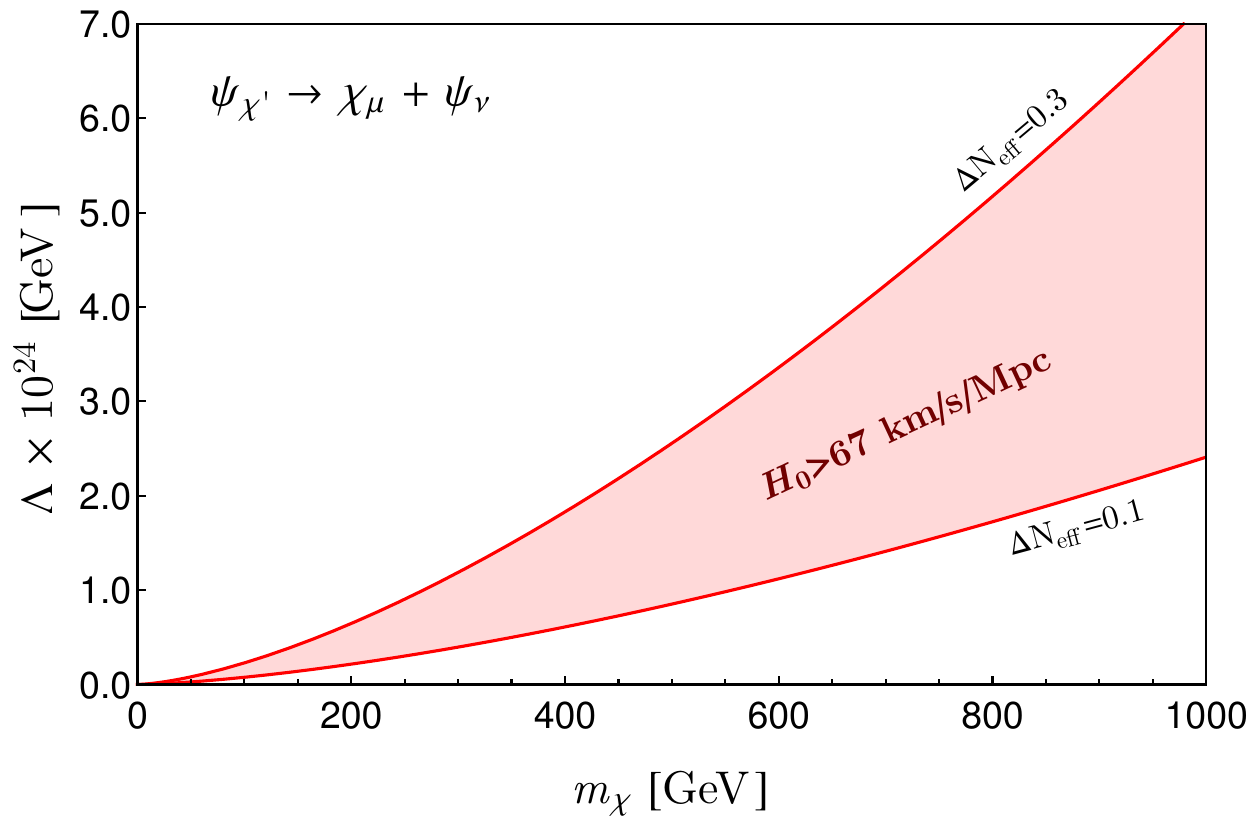}
    \label{fig:plot_Lambda_mchi_mod6}}
    \caption{Region of parameter space that induces that yields $H_0 > 67\, {\rm km s^{-1}Mpc^{-1}}$ and ameliorates the $H_0$ tension for vector dark matter.  In the upper panel, we exhibit the region in the $\Lambda \times m_{\chi^\prime}$ plane, and in the bottom panel $\Lambda \times m_{\chi}$. We have set $f=0.01$ and $m_{\chi^\prime}/m_{\chi} = 10^3$ throughout.}
    \label{fig:plots_casevector}
\end{figure}

Notice that the decay rate for vector dark matter is the largest one compared with the previous cases.  
Plugging this information into Eq.\eqref{eq:deltaN} we get Fig.\ref{fig:MxLambda_case6_FVN} and \fig{fig:plot_Lambda_mchi_mod5}).

Applying Eq. \eqref{eq:gamma_FVN} into Eq. \eqref{eq:deltaN} we obtain,
\begin{equation}
    \Delta N_{eff} = 4.15 \times 10^{-18}\frac{\Lambda}{m_{\chi}}\sqrt{\frac{\mathrm{GeV}}{m_{\chi^\prime}}}f.
    \label{eq:deltaNeff_caseC}
\end{equation}

Comparing this result with the expressions in Eq. \eqref{eq:deltaNeff_caseA} and Eq. \eqref{eq:deltaNeff_caseB}, we conclude the relativistic production of scalar dark matter produces the largest increase in $N_{eff}$. This conclusion is a natural consequence of the decay widths.

\subsection{Discussions}

The effective theories  considered feature $\Gamma \propto m_{\chi^\prime}^3/\Lambda^2$, which implies that $\tau \propto \Lambda^2 / m_{\chi^\prime}^3$, where $m_{\chi^\prime}$ refers to the long-lived heavy particle that decays into a dark matter and neutrino pair. As our relativistic production of dark matter requires $m_{\chi^\prime} \gg m_\chi$, as we increase the dark matter mass we obtain a smaller lifetime and consequently a smaller $\Delta N_{eff}$.  Once we impose $f=0.01$ and $m_{\chi^\prime}/m_\chi$ and $10^2s <\tau <  10^4s$ to be consistent with observational constraints, we find that the relativistic production of dark matter should be mediated by Planck suppressed operators to address the $H_0$ problem if we are interested in dark matter masses around the weak scale.  As $\Delta N_{eff} \propto \frac{\Lambda}{m_\chi}$, If we needed to bring the energy scale down to the Grand Unification scale ($10^{15}$~GeV), we would need dark matter masses around the MeV scale. Having MeV dark matter particles right the right abundance is not trivial, but anyway, the outcome is clear. If one wishes to ameliorate the Hubble tension through the relativistic production of TeV dark matter particles resulted from the decay of long-lived particle, the energy scale dictating the decay rate should be Planck suppressed.

It is important to stress that the other central values of the cosmological parameters obtained from the mechanism employed here are not expected to deviate significantly from the standard $\Lambda$CDM model, neither from any other presently well-established and data-fitted cosmological scenario.

\section{\label{conclusions1} conclusions}

We have investigated the decay of a heavy, long-lived particle into dark matter and neutrinos via the channel $\chi^\prime \rightarrow \chi + \nu$, which can lead to an increase in $N_{eff}$. This occurs because the dark matter particles produced through this decay behave as hot dark matter at the time of matter-radiation equality, effectively mimicking the presence of an additional neutrino species. We have formulated dimension-five effective operators that govern this decay channel for scalar, fermion, and vector dark matter, and we have identified the parameter space where the Hubble tension ($H_0$) can be alleviated through the contribution to $N_{eff}$, considering constraints from Cosmic Microwave Background (CMB), Big Bang Nucleosynthesis (BBN), and structure formation observations. Regardless of the specific nature of the dark matter particle, a significant increase in $N_{eff}$ and a corresponding rise in $H_0$—consistent with local measurements—requires the energy scale controlling the decay process to be near the Planck scale for dark matter masses around the weak scale.

\acknowledgments
The authors thanks Debasish Borah, Oscar Zapata, Yoxara Villamizar, Bertrand Laforge, and Diego Cogollo for discussions. ASJ is supported by Funda\c{c}\~ao de Amparo \`a Pesquisa do Estado de S\~ao Paulo (FAPESP) under the
contract 2023/13126-4. Matheus M. A. Paixão acknowledges the support from CNPQ through grant 151811/2024-5. Dêivid R. da Silva thanks for the support from CNPQ under grant 303699/2023-0. NPN acknowledges the support of CNPq of Brazil under grant PQ-IB 310121/2021-3. 
This work was supported by the Propesq-UFRN Grant 758/2023. The authors acknowledge the use of the IIP cluster ``{\it bulletcluster}". FSQ is supported by Simons Foundation (Award Number:1023171-RC), FAPESP Grant 2018/25225-9, 2021/01089-1, 2023/01197-4, ICTP-SAIFR FAPESP Grants 2021/14335-0, CNPq Grants 307130/2021-5, and ANID-Millennium Science Initiative Program ICN2019\_044.

\appendix

\section{Entropy injection}
\label{secentropy}
As we discuss throughout the text, the non-thermal decay of $\chi'$ into the relativistic dark matter particle $\chi$ and the neutrino adds an extra amount of energy in the form of radiation, what alters the expansion history and changes the value of the Hubble constant. In this appendix we will discuss the entropy injection due to the decay of the long-lived particle.

At the radiation dominated period, the entropy density is given by 
\begin{equation}
    s=\frac{2\pi^2}{45}g_{*s}T^3,
\end{equation}
where $T$ is the photons temperature, while $g_{*s}$ is defined in terms of the bosinic ($g_b$) and fermionic ($g_f$) degrees of freedom as, 
\begin{equation}
    g_{*s}(T) = \sum_b g_b \left( \frac{T_b}{T} \right)^3 + \frac{7}{8} \sum_f g_f\left( \frac{T_f}{T} \right)^3,
    \label{eq:g*s}
\end{equation}
Note that this definition sums over all the relativistic bosons (represented by the index b) and fermions (index f) present in the universe at that moment, with $T_b$ and $T_f$ being the temperatures of each constituent. Therefore, considering the volume dependence with the scale factor $a$, the ratio between the entropy for two different moments is simply,
\begin{equation}
    \frac{S_{\mathrm{f}}}{S_{\mathrm{i}}}=\frac{(g_{*s})_{\mathrm{f}} }{(g_{*s})_{\mathrm{i}}}\left(\frac{a_{\mathrm{f}}T_{\mathrm{f}}}{a_{\mathrm{i}}T_{\mathrm{i}}}\right)^3,
\end{equation}
where f designs the final moment, after the decay, and i represents the initial moment, before the decay. 

Considering just the radiation energy density contribution in equation \eqref{FE},
\begin{equation}
    \rho_{\mathrm{rad}} = \frac{3}{32\pi t^2},
\end{equation}
where we have used the fact that $H^2=1/2t^2$ at radiation-dominated epoch. However, by thermodynamic considerations, we also can express the energy density in terms of the temperature 
\begin{equation}
    \rho_{\mathrm{rad}}=\frac{\pi^2}{30}g_{*}T^2,
\end{equation}
with $g_{*}$ having a similar definition 
\begin{equation}
    g_{*}(T) = \sum_b g_b \left( \frac{T_b}{T} \right)^4 + \frac{7}{8} \sum_f g_f\left( \frac{T_f}{T} \right)^4.
    \label{eq:g*}
\end{equation}
Hence, taking into account that $a\propto t^{1/2}$ at that time,
\begin{equation}
    \frac{S_f}{S_i}=\frac{(g_{*s})_f}{(g_{*s})_i}\left[\frac{(g_{*})_i}{(g_{*})_f}\right]^{3/4}.\label{entropy}
\end{equation}

Within the $\Lambda$CDM model, the entropy remains almost constant between the BBN and the CMB \cite{dodelson2020}. This fact can be understood from the previous equation. At this period we mainly have the contribution of photons and neutrinos to the radiation content. Initially, we have three neutrino families, so $N_{\nu}=3$. Since the photons have two polarizations and because the neutrinos are left-handed, $g_{\gamma}=2$ and $g_{\nu}=1$, respectively, where we need to consider the contributions of the antiparticles as well. As a result, $(g_{*s})_i=3.91$ and $(g_{*})_i=3.36$, in which we account for the fact that $T_{\nu}/T_{\gamma}=(4/11)^{1/3}$. If no decays happen, $(g_{*s})_{i}$ and $(g_{*})_i$ remains constant throughout the evolution, so that no entropy is injected into the system and $\Delta S=0$.

However, the scenario is slightly different when we consider the dark matter decay. The extra amount of radiation present in the universe implies in a final state in which $(g_{*s})_f$ and $(g_{*})_f$ are,
\begin{align}
    (g_{*s})_{f} & = 2 + \frac{7}{4}(3+\Delta N_{eff}) \left( \frac{4}{11} \right), \\
    (g_{*})_{f} & = 2 + \frac{7}{4}(3+\Delta N_{eff}) \left( \frac{4}{11} \right)^{4/3},
\end{align} depending on $\Delta N_{eff}$. Defining the relative entropy variation as $\Delta S/S_i\equiv(S_f-S_i)/S_i$ and using relation \eqref{entropy}, we observe that even with $\Delta N_{eff}$ close to the unity, the injection of entropy corresponds to less than 6\% of the initial entropy. Therefore, we can conclude that the decay does not add a considerable amount of entropy in agreement with CMB observations and BBN. 

\nocite{*}
\bibliography{ref}%

\end{document}